\documentclass[aps,prl,showpacs,twocolumn,superscriptaddress,longbibliography]{revtex4-2}
\usepackage{amsfonts}
\usepackage{amssymb}
\usepackage{amsmath}
\usepackage{graphicx}
\usepackage{epsfig}
\usepackage{subfigure}
\usepackage{appendix}
\usepackage{hyperref}
\usepackage{url}
\hypersetup{hypertex=true,
	colorlinks=true,
	linkcolor=blue,
	urlcolor = blue,
	citecolor=blue}

\begin{document}
	
	\title{Emerging topological characterization in non-equilibrium states of
		quenched Kitaev chains}
	\author{Y. B. Shi}
	\affiliation{School of Physics, Nankai University, Tianjin 300071, China}
	\author{X. Z. Zhang}
	\email{zhangxz@tjnu.edu.cn}
	\affiliation{College of Physics and Materials Science, Tianjin Normal University, Tianjin
		300387, China}
	\author{Z. Song}
	\email{songtc@nankai.edu.cn}
	\affiliation{School of Physics, Nankai University, Tianjin 300071, China}
	
	\begin{abstract}
		Topological characteristics of quantum systems are typically determined by the
			closing of a gap, while the dynamical quantum phase transition
			(DQPT) during quantum real-time evolution has emerged as a nonequilibrium
			analog to the quantum phase transition (QPT). In this paper, we illustrate
			that the system dynamics can be elucidated by considering the precession of a collection of free-pseudo spins under a magnetic field based on the exact
			results of extended Kitaev chains. The topology of the driven Hamiltonian
		is determined by the average winding number of the non-equilibrium state.
		Furthermore, we establish that the singularity of the DQPT arises from two
		perpendicular pseudo-spin vectors associated with the pre- and post-quenched
		Hamiltonians. Moreover, we investigate the distinct behaviors of the dynamic
		pairing order parameter in both topological and non-topological regions.
		These findings offer valuable insights into the non-equilibrium behavior of
		topological superconductors, contributing to the understanding of the
		resilience of topological properties in driven quantum systems.
	\end{abstract}
	
	\maketitle
	
	\affiliation{School of Physics, Nankai University, Tianjin 300071, China} 
	\affiliation{College of Physics and Materials Science, Tianjin Normal University,
		Tianjin 300387, China} 
	\affiliation{School of Physics, Nankai University,
		Tianjin 300071, China}
	
	\textit{Introduction}.-- Quantum theory provides a comprehensive framework
	for describing the equilibrium properties of quantum matter. However, earlier significant theoretical \cite%
		{calabrese2011quantum,heyl2013dynamical,heyl2018dynamical,marino2022dynamical,li2023probing,alba2018entanglement}
		and experimental progress \cite{jurcevic2017direct,zhang2017observation2}
		has greatly contributed to understanding the emergent behavior of isolated quantum
		systems beyond the conventional equilibrium paradigm \cite%
		{monroe2021programmable,blatt2012quantum,schreiber2015observation,bernien2017probing,choi2017observation,wallraff2004circuit,xu2020probing,chang2018colloquium,ye2008quantum,raimond2001manipulating,gring2012relaxation,neyenhuis2017observation,smith2016many,choi2016exploring,zhang2017observation1,zhang2017observation2,jurcevic2017direct}. One approach to studying such non-equilibrium many-body systems is through
	quench dynamics. By using quench techniques, researchers can access novel exotic quantum states
		with energy levels significantly different from those of the ground state. These non-equilibrium states offer insights into the topological and superconducting properties of pre-quenched and
		post-quenched Hamiltonians \cite{shi2022dynamic}. Furthermore, the Loschmidt
		amplitudes
		\begin{equation}
			\mathcal{G}\left( t\right) =\prod_{0<k<\pi } \langle \psi
			(t)\left\vert \psi (0)\right\rangle ,  \label{DQPT}
		\end{equation}
		between the evolved state and initial states exhibit non-analytic behavior over time when the quench crosses a topological phase boundary. This phenomenon is referred to as dynamical quantum phase transition (DQPT) \cite{heyl2015scaling}.
		
		In this work, we investigate the non-equilibrium behavior of the
		general superconducting Kitaev chain with long-range couplings \cite%
		{altland1997nonstandard,kitaev2001unpaired,soori2024majorana,soori2024majorana,decker2024density,malarddetecting,silva2024hybridization,starchl2022relaxation}. The model consists of spinless fermions capable of pairing up to form superconducting Cooper pairs with opposite momenta \cite%
	{vodola2014kitaev,vodola2015long,viyuela2016topological,lepori2017long,bhattacharya2019critical,koffel2012entanglement,hauke2013spread,grass2014trapped,dutta2017probing,vajna2015topological,sedlmayr2018bulk,maslowski2023dynamical,cheraghi2023dynamical,sim2022quench,molignini2017sensing,molignini2018universal,molignini2020edge}. Typically, the topological phase boundaries of a general Kitaev chain are determined by the closing of the gap within the equilibrium paradigm. From the perspective of quantum quench dynamics, we decompose the study of
	non-equilibrium dynamics into an examination of an ensemble of free-pseudo
	spins under a magnetic field. The performance of the evolved state can be
	understood through the average spin precession. The topology of the driven
	Hamiltonian can be deduced from the average winding number of the evolved
	state. We demonstrate that the singularity of the DQPT stems from two
	perpendicular pseudo-spin vectors belonging to the pre- and post-quenched
	Hamiltonians. We also analyze the distinct behaviors of the dynamic
		pairing order parameter in the topological and trivial regions based
		on the non-equilibrium pairing state generated from the vacuum state \cite%
		{shi2022dynamic}, which is defined as 
		\begin{equation}
			\widehat{\zeta}_{k}=-i\left(c_{k}^{\dagger}c_{-k}-c_{k}c_{-k}\right).
		\end{equation}
	In the topological region, the order parameter associated with the pairing
	transition is independent of the chemical potential $\mu$ but decays as $%
	\left\vert \mu\right\vert $ increases.
	
	Our findings provide insights into the non-equilibrium behavior of
	topological superconductors and the interplay between quantum dynamics and
	topology. They have implications for understanding the robustness of
	topological properties in driven quantum systems and exploring novel
	phenomena in quantum information processing and condensed matter physics. 
	\begin{figure}[tbh]
		\centering\includegraphics[width=0.48\textwidth]{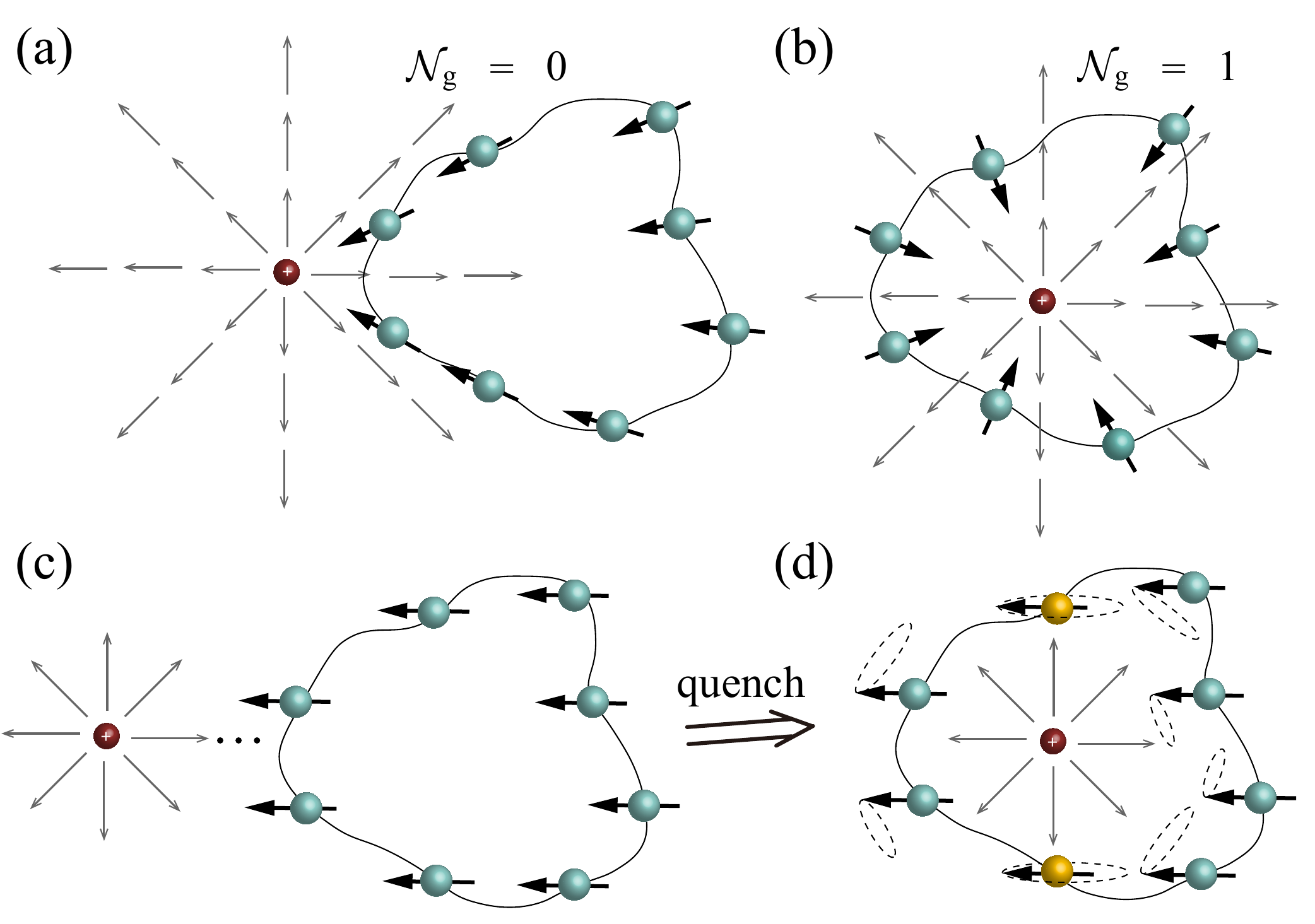}
		\caption{Schematic representation of the ensemble of free pseudo spins under
			a magnetic field. The red solid circle represents the origin and can be
			considered as the monopole. (a)-(b) Different topologies of the Kitaev
			chain. The GS is formed by the direct product of the distinct pseudo-spin
			qubits. The corresponding pseudo-spin vector $\mathbf{r_{\mathrm{g}}}$
			points in the opposite direction of the magnetic field. The topology of the
			system is determined by the winding number of $\mathbf{r_{\mathrm{g}}}$.
			(c)-(d) The quench process employed in this work. The system is initially
			prepared in the topological trivial GS and subsequently quenched to the
			topological region. The yellow solid circle represents the special initial $%
			\mathbf{r}_{\mathrm{I}}\left(k_{c}\right)$, which is perpendicular to the
			local magnetic field of the post-quenched Hamiltonian. The dynamic behavior
			of the non-equilibrium state can be comprehended through the precession of
			each pseudo spin.}
		\label{fig1}
	\end{figure}
		\begin{figure}[tbh]
		\centering\includegraphics[width=0.40\textwidth]{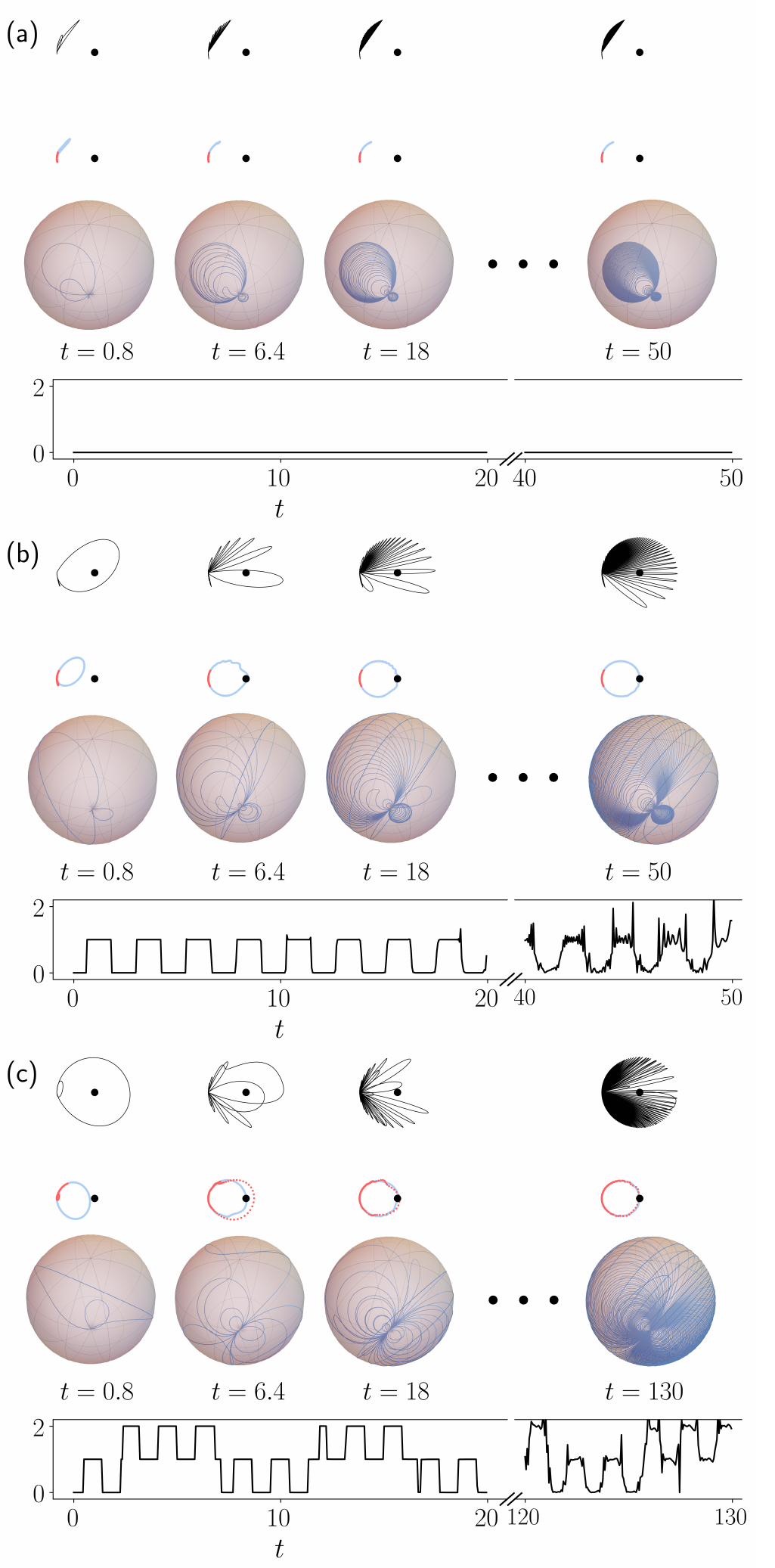}
		\caption{Illustration of the evolution of pseudo-spin vectors after sudden
			quenching. The post-quench Hamiltonians have potentials $\protect\mu $ of
			(a) $4$, (b) $2$, and (c) $-0.5$, with winding numbers $0$, $%
			1 $, and $2$ respectively. The loops of $r_{z}\left( k,t\right) $ and $%
			r_{y}\left( k,t\right) $ are displayed in the first row of each subplot. The
			second row exhibits the plots of the $\overline{r}_{\protect\alpha }\left(
			k,t\right) $. When $t\rightarrow \infty $, the Bloch vector stabilizes, offering valuable insights into the topology of the driven system, as illustrated on the right-hand side. To indicate the ranges, red dots
			represent $[0,{\protect\pi })$ and green dots represent $[{\protect\pi },2%
			\protect\pi )$. Additionally, the third row presents 3D plots of the
			pseudo-spin vector's loop on the Bloch sphere. The fourth row provides a
			characterization of the instantaneous winding number corresponding to the
			instant loops in (a). The remaining parameters are $\Delta _{1}=1$, $\protect%
			\kappa _{1}=1$, $\Delta _{2}=1$, and $\protect\kappa _{2}=2$.}
		\label{fig2}
	\end{figure}

	\textit{Model.--}We examine the following generic Kitaev chain
		\cite{supp,suzuki2019photoinduced}. 
		\begin{equation}
			H=\sum\limits_{n=1}^{M}\sum\limits_{j=1}^{N}(\kappa _{n}c_{j}^{\dag
			}c_{j+n}+\Delta _{n}c_{j}^{\dag }c_{j+n}^{\dag }+\mathrm{H.c.})+\mu
			\sum\limits_{j=1}^{N}\left( 1-2c_{j}^{\dagger }c_{j}\right) , \label{Kitaev}
		\end{equation}%
		where $c_{j}$ denotes the spinless fermion operator at the $j$th
		site with chemical potential $\mu $, which is subject to periodic boundary
		conditions, i.e., $c_{N+n}=c_{n}$. Here $\kappa_{n}$ represents long-range hopping and $\Delta _{n}$ represents the long-range pairing terms. This model has a rich phase diagram and serves as an extended
		one-dimensional mean-field representation of a triplet superconductor. By using the Fourier transformation $%
		c_{j}=\frac{1}{\sqrt{N}}\sum\limits_{k}e^{ikj}c_{k}$ and introducing the pseudo-spin operators 
		\begin{eqnarray}
			s_{k}^{-} &=&\left( s_{k}^{+}\right) ^{\dag }=c_{k}c_{-k},  \notag \\
			s_{k}^{z} &=&\frac{1}{2}\left( c_{k}^{\dag }c_{k}+c_{-k}^{\dag
			}c_{-k}-1\right) .
		\end{eqnarray}%
		The Hamiltonian can be expressed as, 
		\begin{equation}
			H=\sum\limits_{k>0}H_{k}=4\sum\limits_{k>0}\mathbf{B}\left( k\right) \cdot 
			\mathbf{s}_{k}.  \label{H_kitaev}
		\end{equation}
	Here the pseudo-spin operator $s_{k}^{x}=\frac{1}{2}
	\left(s_{k}^{+}+s_{k}^{-}\right)$ and $s_{k}^{y}=\frac{1}{2i}
	\left(s_{k}^{+}-s_{k}^{-}\right)$ satisfy commutation relations of the Lie
	algebra $\left[s_{k}^{\alpha},s_{k^{\prime}}^{\beta}\right]
	=2\delta_{kk^{\prime}}\epsilon_{\alpha\beta\gamma}s_{k^{\prime}}^{\gamma}$,
	with $\epsilon_{\alpha\beta\gamma}$\ Levi-Civita symbol, $%
	\alpha,\beta,\gamma=x,y,z$. Additionally, $\mathbf{B}\left(k\right)=(0,
	\sum_{n=1}^{M}\Delta_{n}\sin\left(nk\right),\sum_{n=1}^{M}\kappa_{n}\cos
	\left(nk\right)-\mu)$ represents an external magnetic field that depends on
	the Bloch momentum ($k$) within the Brillouin zone (BZ). It is worth noting that the pseudo-spin representation of $H$ differs slightly from
	the Pauli matrix representation used for a topological insulator. In this context, the pseudo-spin quantum number is $1/2$ ($0$) in the even-(odd-)parity subspace. In the following we focus on the even-parity subspace which involves quantum number $1/2$ for all $k>0$, aligning with the parity symmetry of the generic Kitaev chain. We choose the $s_{k}^{z}$-basis, denoted as $\{|+,k\rangle =c_{k}^{\dagger }c_{-k}^{\dagger }|0\rangle ,|-,k\rangle
		=|0\rangle \}$, where $|0\rangle $ represents the fermionic vacuum state.
		
		We begin by elucidating how to extract topological information from the ground state via pseudo spins. As evidenced by the fact that $\left[ H_{k},H_{k^{\prime }}\right] =0
		$, we can describe any state $\left\vert \psi \right\rangle
		=\prod_{k}\left\vert \psi \right\rangle _{k}$ within the subspace
		as an ensemble of interaction-free particles,
		each represented by the pseudo-spin vector $\mathbf{r}\left(
		k\right) =(\left\langle s_{k}^{x}\right\rangle ,\left\langle
		s_{k}^{y}\right\rangle ,\left\langle s_{k}^{z}\right\rangle )$, where
		$\left\langle s_{k}^{\alpha }\right\rangle =\left\langle \psi \right\vert
		_{k}s_{k}^{\alpha }\left\vert \psi \right\rangle _{k}$. Consequently, the
		Hamiltonian (\ref{H_kitaev}) can be interpreted as an ensemble of independent spins
		arranged on a loop under the influence of a 2D magnetic field generated by a
		Dirac monopole \cite{dirac1931quantised,barankov2006synchronization}, which has also been demonstrated in the
		previous studies \cite{zhang2017majorana,zhang2015topological}.
	
	Based on this analysis, the ground state (GS) $|G\rangle $ and the phase diagram can be obtained. Each spin constituting the ground state aligns in
	the opposite direction of the $k$-dependent magnetic field, as shown in
	Figs. \ref{fig1}(a) and (b).  Specifically, the vector for the ground state
	$\mathbf{r}_{\mathrm{g}}\left( k\right) $ satisfies 
	\begin{equation}
		\left\langle s_{k}^{\alpha }\right\rangle _{\mathrm{g}}=\left\langle \mathrm{%
			G}\right\vert s_{k}^{\alpha }\left\vert \mathrm{G}\right\rangle =-\frac{%
			B_{\alpha }\left( k\right) }{2B\left( k\right) }.
	\end{equation}%
Consequently, in the thermodynamic limit, the ground state energy density can be described as the integration of the absolute value of the magnetic field, given by $\varepsilon_g = \lim_{N\rightarrow \infty} \frac{E_g}{N} = - \frac{1}{2\pi}\int_0^\pi B(k) \mathrm{d}k$. Simultaneously, $\mathbf{r}_{\mathrm{g}}\left( k\right)$ can construct a closed curve in the $yz$-plane, with the associated winding number, indicating the total counterclockwise revolutions of the curve around a point,
		\begin{equation}
			\mathcal{N}_{\mathrm{g}}= \frac{1}{2\pi }\int\nolimits_{c}\frac{ {r}_{\mathrm{g}}^z \mathrm{d}{r}_{\mathrm{g}}^y-{r}_{\mathrm{g}}^y \mathrm{d}{r}_{\mathrm{g}}^z}{{r}_{\mathrm{g}}^{2}}
		\end{equation}
		This winding number aligns with that derived from the Hamiltonian via $\mathbf{B}(k)$. Notably, the ground state energy density,  which is dependent on the curve configuration, demonstrates nonanalytic behavior when the winding number of the corresponding loop changes. This observation highlights the relationship between the QPT and the geometric topological index that characterizes the phase diagram. In the SM \cite{supp}
		, a comprehensive analysis is provided for various long-range Kitaev models and their associated geometric characteristics.
	Notably, the results presented in this paper exhibit a remarkable degree of generality, extending to arbitrary curves. This observation highlights the dynamic universality inherent in the long-range
		Kitaev model, rendering it applicable across a wide range of system
		configurations.

	\textit{Dynamic Bloch vector}.-- We observe that the Bloch vector denoted
	as $\mathbf{r}_{\mathrm{g}}\left(k\right)$, characterizes topological
	features of the ground state. This concept specifically describes the state
	itself, irrespective of its origin, such as the mother Hamiltonian. Hence, a
	natural question arises: can the Bloch vector detect features arising from
	quantum quenches in the Kitaev model, which differ from merely assessing a static quantity? To address this question, we initially
	focus on the pseudo-spin vector, which initially points to the north pole of
	the Bloch sphere ($\mathbf{r}\left(0\right)=\frac{1}{2}\left(0,0,1\right)$),
	and is subsequently influenced by a magnetic field $\mathbf{B}
	=B\left(\sin\theta\cos\varphi,\sin\theta\sin\varphi,\cos\theta\right)$. The
	time evolution of the pseudo-spin vector can be observed as precession along
	the direction of the magnetic field, which is given by 
	\begin{eqnarray}
		\mathbf{r}\left(t\right) & = & \frac{1}{2}(\sin2Bt\sin\theta\sin\varphi+
		\sin^{2}Bt\sin2\theta\cos\varphi,\text{ }  \notag \\
		& & -\sin2Bt\sin\theta\cos\varphi+\sin^{2}Bt\sin2\theta\sin\varphi,\text{ } 
		\notag \\
		& & 1-\sin^{2}Bt\sin^{2}\theta).
	\end{eqnarray}
	Therefore the average value of the pseudo-spin vector $\mathbf{r}
	\left(t\right)$ over an extended period of time, $\overline{\mathbf{r}}
	\left(t\right)=(\int_{0}^{t}\mathbf{r}\left(t\right)d\tau)/t,$\ can reflect
	the direction of the magnetic field, i.e., $\overline{\mathbf{r}}
	\left(\infty\right)\times\mathbf{B}=0$. Indeed, $\overline{\mathbf{r}}
	\left(\infty\right)$\ corresponds to the projection of $\mathbf{r}
	\left(0\right)$\ onto the direction of the magnetic field.
	
	With this approach in mind, we consider the following protocol \cite{degottardi2011topological,mishra2020disordered}: the initial
	state is the empty state in real space $\left\vert \psi (0)\right\rangle
	=\prod_{\pi >k>0}\left\vert 0\right\rangle _{k}\left\vert 0\right\rangle
	_{-k}=\prod_{l}^{N}\left\vert 0\right\rangle _{N}$. Each pseudo-spin vector
	within the corresponding spin ensemble points toward the $-z$ direction.
	This initial state lacks any information regarding the Hamiltonian. In a
	quantum quench experiment, such a system could be prepared in the ground
	state for infinite potential, i.e., $\mu \rightarrow \infty $ as shown in
	Fig. \ref{fig1}(c). Next, we introduce the post-quench Hamiltonian, $H\left(
	\mu _{0}\right) $ and for any given time $t$, the evolved state
		remains a tensor product state $\left\vert \psi (t)\right\rangle
	=\prod_{\pi >k>0}\left\vert \psi _{k}(t)\right\rangle =\prod_{\pi >k>0}\exp
	(-iH_{k}t)\left\vert \psi _{k}(0)\right\rangle $. A schematic illustration
	of the time evolution is plotted in Fig. \ref{fig1}(d). This indicates that
	every pseudo-spin vector rotates around the magnetic field at that point.
	Consequently, the ensemble $r_{\alpha }\left( k,t\right) =\langle \psi
	(t)|s_{k}^{\alpha }\left\vert \psi (t)\right\rangle $ allows us to extract
	the instantaneous winding number $\mathcal{N}_{t}$ in the $yz$ plane. This
	can be characterized by the last row of each subplot in Fig. \ref{fig2}. The
	time evolution of $\{\mathbf{r}\left( k,t\right) \}$ is presented in the
	third row of Fig. \ref{fig2}. Initially, at $t=0$, it represents a point on
	the Bloch sphere, but for nonzero $t$, it transforms into a curve. As
	expected, the time-dependent curve exhibits high-frequency oscillations for
	large time values. Notably, although the corresponding winding number is not
	constant, it consistently oscillates around a certain value. This
	observation inspires us to average the oscillations by integrating them out 
	\cite{shi2022dynamic}. Over an extended period, the rapidly oscillating
	term becomes negligible, resulting in a steady value given by 
	\begin{eqnarray}
		\overline{r}_{\alpha }^{t}\left( k,t\right) &=&\frac{1}{t}%
		\int_{0}^{t}r_{\alpha }\left( k,\tau \right) d\tau \\
		\overline{r}_{\alpha }\left( k\right) &=&\lim_{t\rightarrow \infty }%
		\overline{r}_{\alpha }\left( k,t\right) =\left\langle s_{k}^{\alpha
		}\right\rangle _{\mathrm{g}}\cos \theta _{k}.
	\end{eqnarray}%
where $\theta _{k}=\cos ^{-1}\left( B_{z}\left( k\right) /B\left(
	k\right) \right) $. The loop of $\{\overline{\mathbf{r}}\left( k,t\right) \}$
	is shown in the second row in Fig. \ref{fig2}. It is evident that the
	corresponding winding numbers of the pseudo-spin vectors $\{\overline{%
		\mathbf{r}}\left( k\right) \}$ and $\{r_{\mathrm{g}}\left( k\right) \}$
	possess the same value. Therefore, the non-equilibrium steady value of $\{%
	\overline{\mathbf{r}}\left( k\right) \}$ reveals the topology of the driven
	Hamiltonian. Indeed, averaging over a finite momentum shell $\left[ k-\Delta
	k/2,k+\Delta k/2\right] $ for the pseudo-spin vector in the limit of large $%
	t $ yields the same result, as mentioned in the SM \cite{supp}. This
	averaging over $k$ helps eliminate the fast oscillation term and provides a
	more robust characterization of the system's properties. 
	
	\begin{figure}[tbh]
		\centering\includegraphics[width=0.49\textwidth]{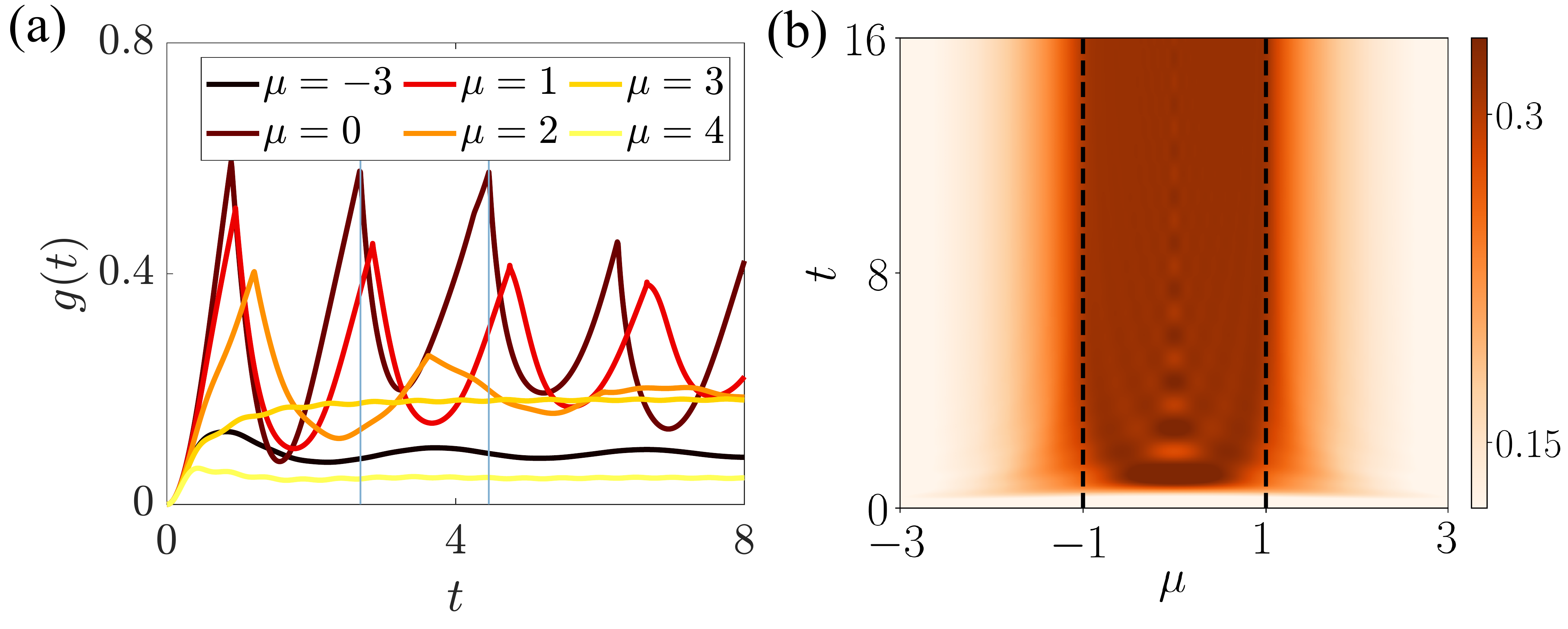}
		\caption{(a) Dynamics of $g(t)$ defined in Eq. (\ref{gt}) following a quench in the generalized Kitaev
			model. Nonanalytic structures are observed when crossing the topological
			phase boundary ($\protect\mu=0,1,2$). The solid blue lines represent the
			time difference between two adjacent nonanalytic behaviors. The numerical
			result of $1.79$ aligns with our analytical findings. The remaining
			parameters are $\Delta_{1}=1$, $\protect\kappa_{1}=1$, $\Delta_{2}=1$, and $%
			\protect\kappa_{2}=2$. (b) Variation of the order parameter $\overline{ 
				\protect\zeta}(t)$ as a function of $t$ and $\protect\mu$. The plot
			demonstrates the tendency of the order parameter to stabilize within a
			plateau in the nontrivial topological region. The black dashed lines
			indicate the analytical behavior of the order parameter at the phase
			boundary.}
		\label{fig3}
	\end{figure}
	
	\textit{Anatomy of the dynamical quantum phase transition}.--DQPT is a
	concept that extends the notion of quantum phase transitions to coherent
	quantum real-time evolution \cite%
	{heyl2013dynamical,heyl2018dynamical,heyl2015scaling,mondal2023finite,ghosh2024corner,mondal2024persistent}%
	. Here we examine the dynamical behavior of two observables to unravel their
	underlying mechanisms based on spin precession.
	
The first observable is the average natural logarithm of the  $\left \vert \mathcal{G}(t) \right \vert ^2 $
		in the large scale limit $N\rightarrow \infty$,
		\begin{equation}
			g\left( t\right) =-\lim_{N\rightarrow \infty }\log  \left \vert \mathcal{G}(t) \right \vert^2 /N, \label{gt}
		\end{equation}%
where $\mathcal{G}(t)$ is defined in Eq. (\ref{DQPT}). $g\left( t\right) $ exhibits nonanalytic behavior when
		the quench process crosses the equilibrium topological phase boundary \cite%
		{heyl2013dynamical}. This phenomenon can be understood through spin
		precession. Let us consider the initial state $\left\vert \psi
	(0)\right\rangle $ which is prepared in the ground state of the Hamiltonian $%
	H$ with a winding number $\mathcal{N}$ of $0$. The corresponding pseudo-spin
	vector is denoted by $\{\mathbf{r}_{\mathrm{I}}\left( k\right) \}$. The $k$%
	-dependent magnetic field $\mathbf{B}_{\mathrm{F}}\left( k\right) $ is
	extracted from the pseudo-spin vector of the ground state of the quenched
	Hamiltonian given by $\mathbf{B}_{\mathrm{F}}\left( k\right) =-\mathbf{r}_{%
		\mathrm{F}}\left( k\right) $ . For any sudden quench across the topological
	phase boundary, there must exist a $k_{c}$ such that $\mathbf{r}_{\mathrm{I}%
	}\left( k_{c}\right) \perp \mathbf{r}_{\mathrm{F}}\left( k_{c}\right) $
	results in $\mathcal{G}_{k_{c}}\left( t_{c}\right) =0$ with critical time $%
	t_{c}=\pi \left( 2n+1\right) /4B\left( k_{c}\right) $, which is represented
	by the yellow solid circle in Fig. \ref{fig1}(d). Fig. \ref{fig3}(a)
	illustrates a quenched process that demonstrates the nonanalytic behavior of 
	$g(t)$. In this scenario, the system is initially prepared on the ground
	state for the initial potential and then quenched to $H(\mu )$ with
	different values of $\mu $. The numerical results of $t_{c}$ correspond well
	with our analysis. The absence of DQPT is evidenced by the smooth and
	analytical evolution of the system, without any abrupt changes or
	singularities, as shown for $\mu =-3,3,4$ in Fig. \ref{fig3}(a).
	
	The pairing is an additional observable of interest. Recent work \cite%
	{shi2022dynamic} has examined two types of order parameters related to local
	pairing in real space and Bardeen-Cooper-Schrieffer (BCS)-like pairing in
	momentum space. A remaining question pertains to why the occurrence of
	topological region for the quench leads to the presence of order parameters
	and the underlying mechanism behind the nonanalytic behavior at the phase
	boundary. This phenomenon can be understood from the perspective of the
	pseudo-spin vector. For the sake of simplicity, we assume the system
	parameters of quenched Hamiltonian as $\kappa_{1}=\delta_{n,1}$, and $%
	\Delta_{n}=\delta_{n,1}\Delta$. Additionally, we recall the expressions for
	BCS-like pairing in momentum space, denoted as 
	\begin{equation}
		\overline{\zeta}(t)=\frac{1}{t}\int_{0}^{t}\frac{\sum_{\pi>k>0}\left\vert
			\left\langle \psi(t)\right\vert \widehat{\zeta}_{k}\left\vert
			\psi(t)\right\rangle \right\vert }{N}d\tau,
	\end{equation}
	with $\widehat{\zeta}_{k}=-i\left(c_{k}^{\dagger}c_{-k}-c_{k}c_{-k}\right)$
	in the context of non-equilibrium dynamics \cite{shi2022dynamic}. Again the
	initial state, $\left\vert \psi(0)\right\rangle =\prod_{l}^{N}\left\vert
	0\right\rangle _{N}$, represents the GS of the pre-quenched topological
	trivial Hamiltonian in the coordinate space when $\mu\rightarrow\infty$.
	Straightforward algebra shows that $\overline{\zeta}(t)=\frac{1}{N}
	\sum_{\pi>k>0}\left\vert F(k)\right\vert \left[1-\frac{\sin\left(4\left\vert 
		\mathbf{B}\left(k\right)\right\vert t\right)}{4\left\vert \mathbf{B}
		\left(k\right)\right\vert t}\right]$, where $F\left(k\right)=\overline{r}
	_{y}\left(k\right)\overline{r}_{z}\left(k\right)/\left\vert \overline{ 
		\mathbf{r}}\left(k\right)\right\vert ^{2}$. We perform the numerical
	simulation in Fig. \ref{fig3}(b). In the thermodynamic limit, it can be
	expressed as follows: $\zeta(\infty)=\frac{1}{4\pi}\int_{0}^{\pi}\left\vert
	\sin2\vartheta_{k}\right\vert dk$, where $\vartheta_{k}=\sin^{-1}\left[ 
	\overline{r}_{y}\left(k\right)/\left\vert \overline{\mathbf{r}}
	\left(k\right)\right\vert \right]$. By considering the contribution of the
	pseudo-spin vector with the same $\sin\vartheta_{k}$, the integral can be
	transformed to $\zeta(\infty)=\frac{1}{2\pi}\int_{0}^{\Theta}\left\vert
	\sin2\vartheta_{k}\right\vert d\varphi_{k}$, with $\tan\vartheta_{k}=\Delta
	\tan\varphi_{k}$. When the post-quenched Hamiltonian resides in the
	nontrivial region, the upper limit $\Theta=\pi/2$, allows $\zeta(\infty)$\
	to attain a value independent of $\mu$. However, in the case of the
	post-quenched Hamiltonian being in the topologically trivial region, $\Theta$
	\ needs to be expressed as a function of $\mu$. Furthermore, the analytic
	behavior of $\zeta(\infty)$ serves as a means to characterize the quantum
	phase transition. For more detailed information, please refer to the SM \cite%
	{supp}.
	
	\textit{Summary and discussion.--}In this letter, we explore the
	non-equilibrium behavior of the Kitaev model with long-range couplings using
	a quantum quench dynamics approach. We investigate an ensemble of
	free-pseudo spins subjected to a magnetic field. The performance of the
	evolved state is analyzed by studying the average spin precession, and the
	topology of the driven Hamiltonian is determined through examination of the
	average winding number of the evolved state. We demonstrate the generation
	of a stable non-equilibrium pairing state starting from an initial vacuum
	state. Furthermore, we establish that the singularity of the DQPT arises
	from two perpendicular pseudo-spin vectors associated with the pre- and
	post-quenched Hamiltonians. We also analyze the distinctive behaviors of the
	dynamic pairing order parameter in both topological and nontopological
	regions. These findings offer valuable insights into the non-equilibrium behavior of topological superconductors, providing implications for
	understanding the resilience of topological properties in driven quantum
	systems. In future research, we aim to extend our investigation to other
	models exhibiting similar characteristics, aiming to broaden our
	understanding of non-equilibrium behavior and uncover potential universality
	and scaling laws.
	
	\acknowledgments We acknowledge the support of the National Natural Science
	Foundation of China (Grants No. 12275193, No. 11975166, and No. 12374461).

\newpage

\begin{widetext}
	\section{Supplemental Material}

\begin{center}
	Y. B. Shi,$^{1}$ X. Z. Zhang,$^{2,\ast }$ and Z. Song$^{1,\dagger }$\\[2pt]
	\textit{$^{1}$School of Physics, Nankai University, Tianjin 300071, China}\\[%
	0pt]
	\textit{$^{2}$College of Physics and Materials Science, Tianjin Normal
		University, Tianjin 300387, China}\\[0pt]
	$^{\ast }$zhangxz@tjnu.edu.cn\\[0pt]
	$^{\dagger }$songtc@nankai.edu.cn \\[0pt]
\end{center}

\setcounter{equation}{0} \renewcommand{\theequation}{S\arabic{equation}} %
\setcounter{figure}{0} \renewcommand{\thefigure}{S\arabic{figure}} %
\setcounter{secnumdepth}{3}

In this Supplemental Material, we present \ref{A}. Configuration of pseudo
spin and topology; \ref{B}. Derivation of dynamics of the Bloch vector and
dynamical topological characterization; and \ref{C}. Derivation of the order
parameter $\overline{\zeta }$.

\subsection{Configuration of pseudo spin and topology}

\label{A} In the main text, we present the extensive topological properties
of the long-range Kitaev model through its geometric characterization.
Within this subsection, we explicitly demonstrate this phenomenon by
examining specific Hamiltonians in coordinate space. The
Hamiltonian of the generalized Kitaev chain is the same as Eq. (3) in the
main text
\begin{equation}
	H=\sum\limits_{n=1}^{M}\sum\limits_{j=1}^{N}(\kappa _{n}c_{j}^{\dag
	}c_{j+n}+\Delta _{n}c_{j}^{\dag }c_{j+n}^{\dag }+\mathrm{H.c.})+\mu
	\sum\limits_{j=1}^{N}\left( 1-2c_{j}^{\dagger }c_{j}\right) ,
\end{equation}%
where $c_{j}$ represents the spinless fermion operator at the $j$th site
with chemical potential $\mu $, and it obeys periodic boundary conditions,
i.e., $c_{N+n}=c_{n}$. This model serves as an extended one-dimensional
mean-field representation of a triplet superconductor, incorporating
long-range hopping and pairing terms with strengths $\kappa _{n}$ and $%
\Delta _{n}$, respectively. In the limit of large $N$ with $M\ll N$, the
Hamiltonian can be mapped to a one-dimensional quantum spin model \cite%
{suzuki2019photoinduced} using the conventional Jordan-Wigner
transformation, 
\begin{equation}
	c_{j}=\prod_{i=1}^{j-1}\left( -\sigma _{i}^{z}\right) \sigma
	^{-},c_{j}^{\dagger }=\prod_{i=1}^{j-1}\left( -\sigma _{i}^{z}\right) \sigma
	^{+}.
\end{equation}%
By employing the Fourier transformation, 
\begin{equation}
	c_{j}=\frac{1}{\sqrt{N}}\sum\limits_{k}e^{ikj}c_{k},  \label{FT}
\end{equation}%
the total Hamiltonian $H$ can be block diagonalized as 
\begin{equation}
	H=\sum\limits_{k>0}H_{k},
\end{equation}%
satisfying $\left[ H_{k},H_{k^{\prime }}\right] =0$ and $H_{k}$ is
given by, 
\begin{equation}
	H_{k}=-2i\sum_{n=1}^{M}\Delta _{n}\sin (nk)c_{-k}^{\dagger }c_{k}^{\dagger
	}+ \mathrm{h.c.}+2\left( \sum_{n=1}^{M}\kappa _{n}\cos (nk)-\mu \right)
	\left( c_{k}^{\dag }c_{k}+c_{-k}^{\dag }c_{-k}\right) -2\mu .
\end{equation}
When we introduce the pseudo-spin operators, 
\begin{eqnarray}
	s_{k}^{-} &=&\left( s_{k}^{+}\right) ^{\dag }=c_{k}c_{-k},  \notag \\
	s_{k}^{z} &=&\frac{1}{2}\left( c_{k}^{\dag }c_{k}+c_{-k}^{\dag
	}c_{-k}-1\right) ,  \label{s_k}
\end{eqnarray}
the Hamiltonian $H_{k}$ can be expressed as 
\begin{eqnarray}
	H_{k} &=&4\left( \sum_{n=1}^{M}\kappa _{n}\cos (nk)-\mu \right)
	s_{k}^{z}-2i\sum_{n=1}^{M}\Delta _{n}\sin
	(nk)s_{k}^{+}+2i\sum_{n=1}^{M}\Delta _{n}\sin (nk)s_{k}^{-}  \label{H_k} \\
	&=&4\mathbf{B}\left( k\right) \cdot \mathbf{s}_{k},
\end{eqnarray}%
with the field $B\left( k\right) =(0,\sum_{n=1}^{M}\Delta _{n}\sin
(nk),\sum_{n=1}^{M}\kappa _{n}\cos (nk)-\mu )$. The pseudo-spin operators%
$s_{k}^{x}=\frac{1}{2}(s_{k}^{+}+s_{k}^{-})$ and $s_{k}^{y}=\frac{ 1}{2i}%
(s_{k}^{+}-s_{k}^{-})$ follow the commutation relations of the Lie algebra $%
\left[ s_{k}^{\alpha },s_{k^{\prime }}^{\beta }\right] =2\delta _{kk^{\prime
}}\epsilon _{\alpha \beta \gamma }s_{k^{\prime }}^{\gamma }$, where $%
\epsilon _{\alpha \beta \gamma }$ denotes the Levi-Civita symbol and $\alpha 
$, $\beta $, and $\gamma $ take on values of $x$, $y$, and $z$.

This conclusion applies to the generalized model at hand, which corresponds
to a loop traced by the parametric equation $\mathbf{B}\left(k\right)$. In
general, the winding number of a closed curve in the auxiliary $yz$-plane
around the origin is defined as $\mathcal{N}=\frac{1}{2\pi }%
\int\nolimits_{c} \frac{1}{r^{2}}\left( z\mathrm{d}y-y\mathrm{d}z\right) $,
an integer representing the total number of times the curve travels
anticlockwise around the origin. The connection between the quantum phase
transition and the switch of the topological quantity has been established
by considering $\mathbf{r}\left( k\right) =\mathbf{B}\left( k\right) $ \cite%
{zhang2017majorana}.

Alternatively, one can introduce a vector $\mathbf{r}_{\mathrm{g}}\left(
k\right) =(\left\langle s_{k}^{x}\right\rangle _{\mathrm{g}},\left\langle
s_{k}^{y}\right\rangle _{\mathrm{g}},\left\langle s_{k}^{z}\right\rangle _{%
	\mathrm{g}})$ with 
\begin{equation}
	\left\langle s_{k}^{\alpha }\right\rangle _{\mathrm{g}}=\left\langle \mathrm{%
		G}\right\vert s_{k}^{\alpha }\left\vert \mathrm{G}\right\rangle =-\frac{%
		B_{\alpha }\left( k\right) }{2B\left( k\right) }.  \label{s_k g}
\end{equation}%
The corresponding winding number $\mathcal{N}_{\mathrm{g}}$ can be obtained
by considering $\mathbf{r}\left( k\right) =\left( \left\langle
s_{k}^{x}\right\rangle _{\mathrm{g}},\left\langle s_{k}^{y}\right\rangle _{%
	\mathrm{g}},\left\langle s_{k}^{z}\right\rangle _{\mathrm{g}}\right) $.
Notably, both vectors $\mathbf{r}_{\mathrm{g}}\left( k\right) $ and $\mathbf{%
	\ B}\left( k\right) $ share the same winding number. In parallel, $r_{%
	\mathrm{g}}\left( k\right) =|\mathbf{r}_{\mathrm{g}}\left( k\right) |$ plays
a similar role to $B\left( k\right) $ in characterizing the occurrence of a
quantum phase transition.

Considering the nonzero parameters as $\kappa _{1}=1$\ and $\Delta _{1}=1$,
the vector $\mathbf{r}_{\mathrm{g}}\left( k\right) =\left( 0,\frac{\sin k}{2 
	\sqrt{1-2\mu \cos k+\mu ^{2}}},\frac{\cos k-\mu }{2\sqrt{1-2\mu \cos k+\mu
		^{2}}}\right) $ represents a circle. The winding numbers associated with it
are $0$\ and $1$, depending on the value of $\mu $. Similarly, when $\kappa
_{1}=1$, $\kappa _{2}=2$, $\Delta _{1}=1$, and $\Delta _{2}=1$, the vector $%
\mathbf{r}_{\mathrm{g}}\left( k\right) =\left( 0,\frac{\sin k+\sin 2k}{2| 
	\mathbf{B}|},\frac{\cos k+2\cos 2k-\mu }{2|\mathbf{B}|}\right) $ represents
a trefoil. The trefoil has winding numbers of $0$, $1$, and $2$. Here, $| 
\mathbf{B}|=\sqrt{\mu ^{2}-2(2\cos 2k+\cos k)\mu +4\cos ^{3}k+3\cos ^{2}2k+2}
$. Figs. \ref{figS1}(a) and (b) illustrate two such typical cases, which
will be also employed to demonstrate the subsequent quench dynamics. It is
worth pointing out that the correspondence between the GS and free spin
subjected to the magnetic field is held for any given geometric curve. For
diversity, we give two complex examples in Figs. \ref{figS1}(c) and (d).

\begin{figure}[tbh]
	\centering\includegraphics[width=0.90\textwidth]{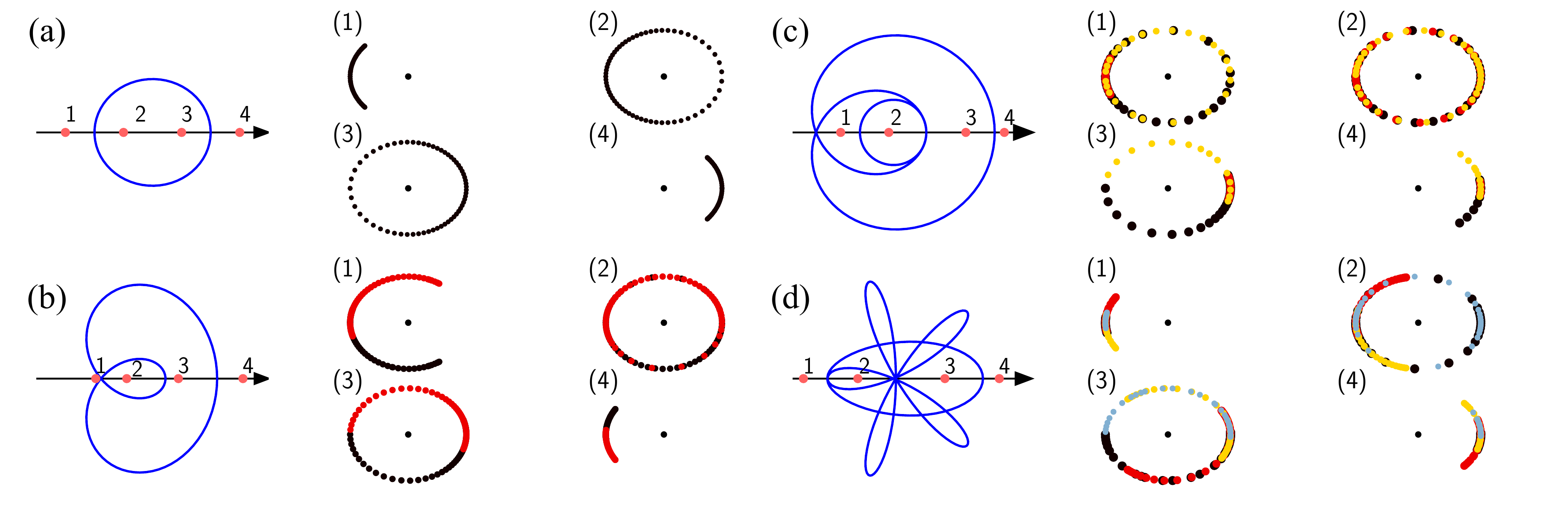} 
	\caption{Geometric representation of the GS of the long-range Kitaev model
		in the parametric space. The left panel of each figure illustrates the curve
		of the GS, which is determined by Eq. (\protect\ref{s_k g}). The red circle
		represents the origin of the parametric space, which determines the topology
		of the Hamiltonian. The right panel shows the curve when the origin is
		determined. The corresponding parameters are as follows: (a) $\protect\kappa 		_{1}=1$\ and $\Delta _{1}=1$. The black points on the right panel indicate
		the range of momentum $k$\ from $(0,2\protect\pi ]$. The locations of the
		origins defined by the red circles are $\protect\mu =-1.5$, $-0.5$, $0.5$,
		and $1.5$, respectively. (b) $\protect\kappa _{1}=1$, $\protect\kappa _{2}=2$		, $\Delta _{1}=1$, and $\Delta _{2}=1$. The black and red points on the
		right panel represent the range of momentum $k$\ from $(0,\protect\pi ]$\
		and $(\protect\pi ,2\protect\pi ]$, respectively. The origins of four
		typical cases are determined by $\protect\mu =-1.7$, $-0.5$, $1.5$, and $4$,
		respectively. (c) $\protect\kappa _{1}=1$, $\protect\kappa _{2}=4$, $\protect		\kappa _{3}=6$, $\Delta _{1}=1$, $\Delta _{2}=2$, and $\Delta _{3}=3$. The
		black, red, and yellow points indicate the range of momentum $k$\ from $(0,		\frac{2}{3}\protect\pi ]$, $(\frac{2}{3}\protect\pi ,\frac{4}{3}\protect\pi ]
		$, and $(\frac{4}{3}\protect\pi ,2\protect\pi ]$, respectively. The
		corresponding $\protect\mu $\ values are $-5$, $0$, $8$, and $12$,
		respectively. (d) $\protect\kappa _{1}=1$, $\protect\kappa _{2}=-2$, $		\protect\kappa _{3}=7$, $\protect\kappa _{4}=9$, $\Delta _{1}=1$, $\Delta
		_{2}=2$, $\Delta _{3}=-3$, and $\Delta _{4}=4$. The black, red, yellow, and
		blue points represent the range of momentum $k$\ from $(0,\frac{1}{2}\protect		\pi ]$, $(\frac{1}{2}\protect\pi ,\protect\pi ]$, $(\protect\pi ,\frac{3}{2}		\protect\pi ]$, and $(\frac{3}{2}\protect\pi ,2\protect\pi ]$, respectively.
		The value of $\protect\mu $\ is $-18$, $-8$, $8$, and $18$, respectively.
		The different colors highlight the larger winding number of the curve.} %
	\label{figS1}
\end{figure}

\subsection{Derivation of dynamics of the Bloch vector and dynamical
	topological characterization}

\label{B} In the previous section, we have demonstrated the utility of the
Bloch vector $\mathbf{r}_{\mathrm{g}}\left( k\right) $ in characterizing the
quantum phase diagram and topological properties of the ground state.
Importantly, this concept solely focuses on the state itself, disregarding
its origin, which is the mother Hamiltonian. Generally, it provides insights
into the characteristics of the associated Hamiltonian, especially when the
state corresponds to the ground state. Consequently, an important question
arises: Can the Bloch vector extracted from a time-evolved state effectively
capture the features of the driven Hamiltonian? We provide an affirmative
answer. In this section, we present two possible methods for detecting the
pseudo-spin vectors. First, we introduce a time-dependent Bloch-like vector
denoted as 
\begin{equation}
	\left\langle s_{k}^{\alpha }\right\rangle _{t}=\left\langle \psi
	(t)\right\vert s_{k}^{\alpha }\left\vert \psi (t)\right\rangle ,
\end{equation}
which is defined as the expectation value of the pseudo-spin operator $%
s_{k}^{\alpha }$ for the evolved state $\left\vert \psi (t)\right\rangle
=e^{-iHt}\left\vert \psi (0)\right\rangle $ $=\prod_{\pi
	>k>0}e^{-iH_{k}t}\left\vert 0\right\rangle _{k}\left\vert 0\right\rangle
_{-k}$. This vector represents a well-defined set of values and the
instantaneous winding number $\mathcal{N}_{t}$\ can be obtained by
considering the components $\{\mathbf{r}\left( k\right) =(\left\langle
s_{k}^{x}\right\rangle _{t},\left\langle s_{k}^{y}\right\rangle
_{t},\left\langle s_{k}^{z}\right\rangle _{t})\}$. In Fig. 2 of the main
text, the time-dependent curve exhibits high-frequency oscillations for
large time values. It is worth noting that although the corresponding
winding number is not constant, it consistently oscillates around a specific
value. This observation motivates us to perform an averaging procedure to
eliminate the oscillations. In the following, we will explore two different
approaches for averaging: (i) averaging over a long time interval $t$ and
(ii) averaging over a finite momentum shell $\left[ k-\Delta k/2,k+\Delta
k/2 \right] $.

First, taking the average over a period of time $t$, we obtain a $t$
-dependent vector denoted as 
\begin{equation}
	\overline{r}_{\alpha }^{t}\left( k,t\right) =\frac{1}{t}\int_{0}^{t}\left
	\langle s_{k}^{\alpha }\right\rangle _{\tau }\mathrm{d}\tau .
	\label{r_average}
\end{equation}
A straightforward derivation yields the following expression for the state
vector $\left\vert \psi (t)\right\rangle $ 
\begin{equation}
	\left\vert \psi (t)\right\rangle =\left[ \cos \left( 2\left\vert \mathbf{B}
	\right\vert t\right) +i\frac{B_{z}}{\left\vert \mathbf{B}\right\vert }\sin
	\left( 2\left\vert \mathbf{B}\right\vert t\right) \right] \left\vert
	0\right\rangle _{k}\left\vert 0\right\rangle _{-k}-\frac{B_{y}}{\left\vert 
		\mathbf{B}\right\vert }\sin \left( 2\left\vert \mathbf{B}\right\vert
	t\right) \left\vert 1\right\rangle _{k}\left\vert 1\right\rangle _{-k}.
\end{equation}
Based on the evolved state, we express the set of pseudo-spin vectors at
each time as follows 
\begin{eqnarray}
	\left\langle s_{k}^{x}\right\rangle _{t} &=&-\frac{B_{y}}{\left\vert \mathbf{%
			\ \ B}\right\vert }\sin \left( 4\left\vert \mathbf{B}\right\vert t\right) , 
	\notag \\
	\left\langle s_{k}^{y}\right\rangle _{t} &=&-\frac{B_{y}B_{z}}{\left\vert 
		\mathbf{B}\right\vert ^{2}}\sin ^{2}\left( 2\left\vert \mathbf{B}\right\vert
	t\right) ,  \notag \\
	\left\langle s_{k}^{z}\right\rangle _{t} &=&\frac{B_{y}^{2}}{\left\vert 
		\mathbf{B}\right\vert ^{2}}\sin ^{2}\left( 2\left\vert \mathbf{B}\right\vert
	t\right) -\frac{1}{2}.
\end{eqnarray}
By substituting these equations into Eq. (\ref{r_average}), we readily
obtain the expressions for $\overline{r}_{{\alpha }}^{t}(k,t)$, i.e., 
\begin{eqnarray}
	\overline{r}_{x}^{t}\left( k,t\right) &=&\frac{B_{y}}{\left\vert \mathbf{B}
		\right\vert }\frac{\cos \left( 4\left\vert \mathbf{B}\right\vert t\right) -1 
	}{4\left\vert \mathbf{B}\right\vert t},  \notag \\
	\overline{r}_{y}^{t}\left( k,t\right) &=&\frac{B_{y}B_{z}\sin \left(
		4\left\vert \mathbf{B}\right\vert t\right) }{8\left\vert \mathbf{B}
		\right\vert ^{3}t}-\frac{B_{y}B_{z}}{2\left\vert \mathbf{B}\right\vert ^{2}},
	\notag \\
	\overline{r}_{z}^{t}\left( k,t\right) &=&-\frac{B_{y}^{2}\sin \left(
		4\left\vert \mathbf{B}\right\vert t\right) }{8\left\vert \mathbf{B}
		\right\vert ^{3}t}-\frac{B_{z}^{2}}{2\left\vert \mathbf{B}\right\vert ^{2}}.
	\label{r_t}
\end{eqnarray}
In the limit as $t\rightarrow \infty $, we have 
\begin{equation}
	\overline{r}_{\alpha }^{t}\left( k\right) =\lim_{t\rightarrow \infty } 
	\overline{r}_{\alpha }^{t}\left( k,t\right) =\left\langle s_{k}^{\alpha
	}\right\rangle _{\mathrm{g}}\cos \theta _{k}.  \label{A_T}
\end{equation}
In this context, $\theta _{k}=\cos ^{-1}\left( \frac{B_{z}(k)}{B(k)}\right) $
represents the angle between the initial Bloch vector and the magnetic field
dependent on $k$. Moreover, the winding numbers of the two vectors$\ 
\overline{r}_{\alpha }(k)$ and $\langle s_{k}^{\alpha }\rangle _{\mathrm{g}}$
possess the same value.

Next, we average each pseudo-spin vector over a finite momentum shell $\left[
k-\Delta k/2,k+\Delta k/2\right] $, given by 
\begin{equation}
	\overline{r}_{\alpha }^{k}\left( k_{0},t\right) =\frac{1}{2\Delta k}
	\int_{k_{0}-\Delta k}^{k_{0}+\Delta k}\left\langle s_{k}^{\alpha
	}\right\rangle _{t}dk\text{,}
\end{equation}
where $k\gg \Delta k\gg dk$. In the limit of large time $t$, 
\begin{eqnarray}
	\overline{r}_{x}^{k}\left( k_{0},t\right) &\approx &\frac{B_{y}}{8\Delta
		kt\left\vert \mathbf{B}\right\vert }\left[ \cos \left( 4\left\vert \mathbf{B}
	\right\vert \left( k_{0}+\Delta k\right) t\right) -\cos \left( 4\left\vert 
	\mathbf{B}\right\vert \left( k_{0}-\Delta k\right) t\right) \right] \frac{dk 
	}{d\left\vert \mathbf{B}\right\vert }|_{k=k_{0}},  \notag \\
	\overline{r}_{x}^{k}\left( k_{0},t\right) &\approx &\frac{B_{y}B_{z}}{
		16\left\vert \mathbf{B}\right\vert ^{2}\Delta kt}\left[ \sin \left(
	4\left\vert \mathbf{B}\right\vert \left( k_{0}+\Delta k\right) t\right)
	-\sin \left( 4\left\vert \mathbf{B}\right\vert \left( k_{0}-\Delta k\right)
	t\right) \right] \frac{dk}{d\left\vert \mathbf{B}\right\vert }|_{k=k_{0}}- 
	\frac{B_{y}B_{z}}{2\left\vert \mathbf{B}\right\vert ^{2}},  \notag \\
	\overline{r}_{x}^{k}\left( k_{0},t\right) &\approx &-\frac{B_{y}^{2}}{
		16\left\vert \mathbf{B}\right\vert ^{2}\Delta kt}\left[ \sin \left(
	4\left\vert \mathbf{B}\right\vert \left( k_{0}+\Delta k\right) t\right)
	-\sin \left( 4\left\vert \mathbf{B}\right\vert \left( k_{0}-\Delta k\right)
	t\right) \right] \frac{dk}{d\left\vert \mathbf{B}\right\vert }|_{k=k_{0}}- 
	\frac{B_{z}^{2}}{2\left\vert \mathbf{B}\right\vert ^{2}}.  \label{r_k}
\end{eqnarray}
Neglecting the high-frequency oscillating term yields the same result as $%
\overline{r}_{{\alpha }}^{t}\left( k,t\right) $, 
\begin{equation}
	\overline{r}_{\alpha }^{k}\left( k\right) =\lim_{T\rightarrow \infty } 
	\overline{r}_{\alpha }^{k}\left( k,T\right) =\left\langle s_{k}^{\alpha
	}\right\rangle _{\mathrm{g}}\cos \theta _{k}.  \label{A_K}
\end{equation}
We consider the value of $\Delta k$\ is proportional to $1/N$\ and the value
of $dk$\ equals $2\pi /N$\ in the numerical simulation. In Fig. \ref{figS2},
we present the ensembles $\left\{ \overline{r}_{\alpha }^{t}\left(
k,t\right) \right\} $ and $\left\{ \overline{r}_{\alpha }^{k}\left(
k,t\right) \right\} $ at different times. The collection of curves depicted
in Fig. \ref{figS2} demonstrates that both of these distinct averaging
approaches serve a common purpose and effectively reveal the topology of the
driven Hamiltonian, aligning with the analytical results presented in Eqs. ( %
\ref{A_T}) and (\ref{A_K}). At the same time, employing longer time
intervals for averaging, denoted as $t$, leads to faster stabilization. The
difference in the time scales at which the two methods converge to stability
is influenced by the parameter $\Delta k$, as evident in Eqs. (\ref{r_t})
and (\ref{r_k}). Moreover, this approach has the potential to facilitate
experimental measurements by providing a simplified and accessible method
for evaluating the order parameter.

Next, we consider two special conditions, one with non-uniform on-site potential and the other with non-uniform coupling strengths. The numerical simulation for the first condition is shown
in Fig. \ref{figS3}. In Fig. \ref{figS3}(a) and \ref{figS3}(b), we use the
same parameters as in Fig. \ref{figS2}, but with on-site disorder within the range $\left[-\delta,\delta \right]$. We choose
different values of $\mu$ to consider different systems whose ground
states have different winding numbers. In Fig. \ref{figS3}(c), we also plot
the time-averaged $\left\{ \overline{r}_{\alpha }^{t}\left( k,t\right)
\right\}$ for the extended version of the system in Fig. \ref{figS1}(c) with
disorder. For the second condition, we consider the inverse-square
decaying coupling strengths, i.e., 
\begin{equation*}
	\kappa _{n}=\frac{1}{n^{2}},\Delta _{1}=1.
\end{equation*}
The numerical simulation is performed in Fig. \ref{figS4}. They all demonstrate
the robustness and universality of our approach.

\begin{figure}[tbh]
	\centering\includegraphics[width=0.90\textwidth]{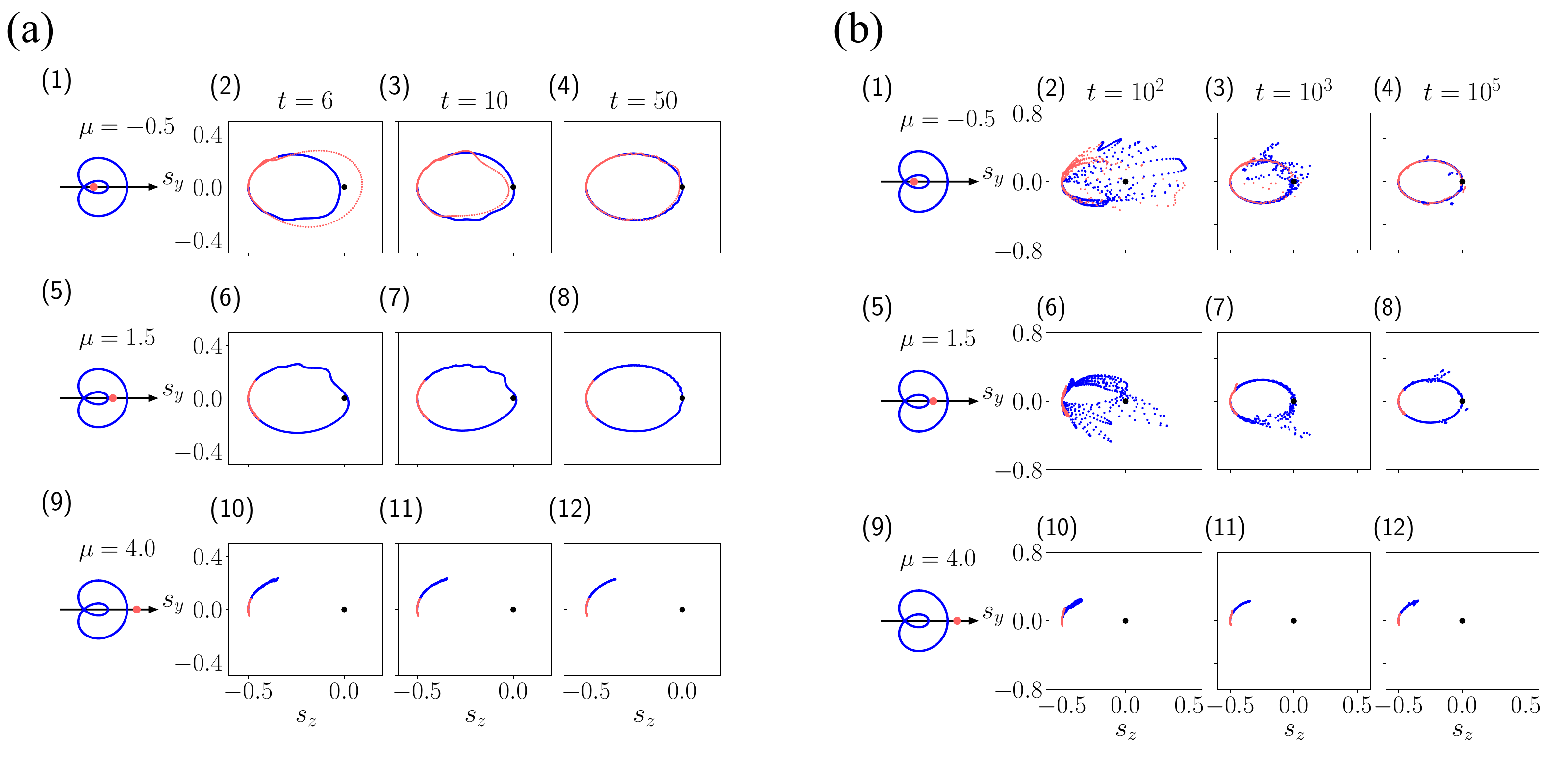} 
	\caption{Plots of (a) time-averaged\ $\left\{ \overline{r}_{\protect\alpha 		}^{t}\left( k,t\right) \right\}$ and (b) momentum-averaged $\left\{ 
		\overline{r}_{\protect\alpha }^{k}\left( k,t\right) \right\} $\ at different
		times. The left panel of each subfigure illustrates the considered system.
		To indicate the ranges, blue dots represent $[0,{\protect\pi })$ and red
		dots represent $[{\protect\pi },2\protect\pi )$. Other parameters are set as
		follows: $N=10^{6}$, $\Delta k=\frac{\protect\pi }{10^{3}}$, $\protect\kappa 		_{1}=1$, $\protect\kappa _{2}=2$, $\Delta _{1}=1$, and $\Delta _{2}=1$.}
	\label{figS2}
\end{figure}

\begin{figure*}[tbh]
	\centering\includegraphics[width=0.90\textwidth]{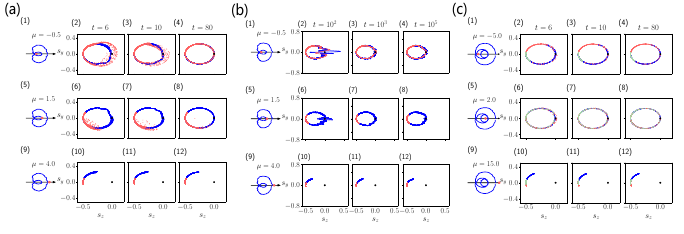}  
	\caption{Plots of (a) time-averaged\ $\left\{ \overline{r}_{\protect\alpha %
		}^{t}\left( k,t\right) \right\} $\ and (b) momentum-averaged  $\left\{ 
		\overline{r}_{\protect\alpha }^{k}\left( k,t\right) \right\} $ \ at
		different times for the non-uniform $\protect\mu $. The left panel of each
		subfigure illustrates the considered system. To indicate the ranges, blue
		dots represent $[0,\protect\pi )$\ and red dots represent $[\protect\pi ,2%
		\protect\pi )$. Other parameters are set as follows: $N=10^{6}$, $\Delta k=%
		\frac{\protect\pi }{10^{3}}$, $\protect\kappa _{1}=1$, $\protect\kappa _{2}=2
		$ , $\Delta _{1}=1$, $\Delta _{2}=1$, and $\protect\delta =0.5$. (c) $%
		\left\{ \overline{r}_{\protect\alpha }^{t}\left( k,t\right) \right\} $\ for
		the extended Kitaev chain. To indicate the ranges, blue
		dots represent $[0,2\protect\pi/3 )$, red dots represent $[2\protect\pi/3 ,4
		\protect\pi/3 )$ and cyan dots represent $[4\protect\pi/3 ,2\protect\pi)$. Other parameters are set as follows: $\protect\kappa _{1}=1$, $\protect\kappa _{2}=4$, $\protect\kappa _{3}=6$, $\Delta
		_{1}=1$, $\Delta _{2}=2$, $\Delta _{3}=3$, and $\protect\delta =0.5$.} 
	\label{figS3}
\end{figure*}

\begin{figure}[tbh]
	\centering\includegraphics[width=0.90%
	\textwidth]{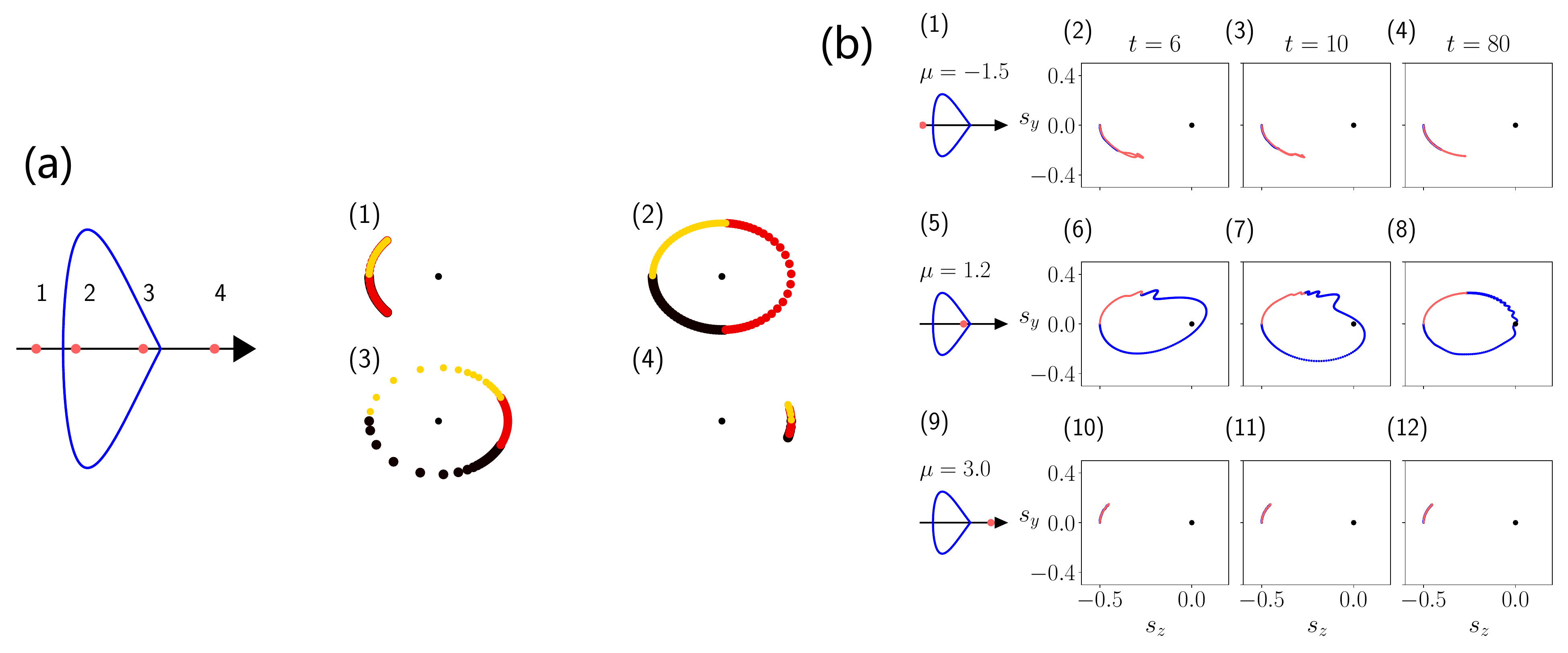}  
	\caption{The description of topological property for the inverse-square
		decay. (a) Geometric representation for the topological property in the
		parametric space. The left panel illustrates the curve of the magnetic
		field, $\mathbf{B}(k)$. The red dots represent the origin of the parametric space, 
		$\mu =-1.5,-0.5,1.2,$and $3$, which determine the topology of the Hamiltonian.
		The right panel shows the curves of the pseudo-spin vector of the GS. (b)
		Time-averaged pseudo-spin $\left\{ \overline{r}_{\protect\alpha }^{t}\left(
		k,t\right) \right\} $.} 
	\label{figS4}
\end{figure}

\subsection{Derivation of the order parameter $\overline{\protect\zeta }$}

\label{C} The consistent outcome of the two distinct averaging approaches
after an extended period in the previous section prompts us to consider
whether the instantaneous order parameter, 
\begin{equation}
	\zeta (t)=\frac{\sum_{\pi >k>0}\left\vert \left\langle \psi (t)\right\vert 
		\widehat{\zeta }_{k}\left\vert \psi (t)\right\rangle \right\vert }{N},
\end{equation}
will eventually show the same expression as the time-averaged order
parameter $\overline{\zeta }(t)$? To answer this question we decompose $%
\zeta \left( t\right) $\ into two parts,

\begin{equation}
	\zeta \left( t\right) =\zeta \left( \infty \right) -\delta \zeta \left(
	t\right),
\end{equation}
where 
\begin{eqnarray}
	\zeta \left( \infty \right) &=&\frac{1}{2\pi }\int_{0}^{\pi }\left\vert
	F\left( k\right) \right\vert \mathrm{d}k, \\
	\delta \zeta \left( t\right) &=&\frac{1}{2\pi }\int_{0}^{\pi }\left\vert
	F\left( k\right) \right\vert \cos \left( 4\left\vert \mathbf{B}\left(
	k\right) \right\vert t\right) \mathrm{d}k.  \label{dnt}
\end{eqnarray}
The $\zeta \left( \infty \right) $\ means the stationary part of $\zeta (t)$
\ and $\delta \zeta \left( t\right) $\ means the oscillatory part. After
evaluating through the saddle\ point integration in the continuum limit, we
will find that $\delta \zeta \left( t\right) $\ behaves as a damping
oscillation\ and that the decay rate depends on the location $k_{0}$\ of the
extrema $\left\vert \mathbf{B}\left( k\right) \right\vert $. (i) In the case
where $k_{0}\neq 0,\pi $, the quasi frequency is approximately $\omega
\approx 2\left\vert \mathbf{B}\left( k_{0}\right) \right\vert $, and the
decay of $\delta \zeta \left( t\right) $ follows a power law with $\sim
t^{-0.5}$. (ii) In the case where $k_{0}=0,\pi $, the quasi frequency is
approximately $\omega \approx 2\left\vert \mathbf{B}\left( k_{0}=0\right)
\right\vert $, and the decay of $\delta \zeta \left( t\right) $ follows a
power law with $\sim t^{-1}$.\ These findings are consistent with the plots
presented in Fig. \ref{figS5}(a) and (b). In conclusion, we obtain 
\begin{equation}
	\overline{\zeta }\left( t\rightarrow \infty \right) =\zeta \left(
	t\rightarrow \infty \right) .
\end{equation}
This means that $\zeta $ could play the same role as $\overline{\zeta}$.

In the following, we focus on the analytic form of $\overline{\zeta }$.
Although the explicit form has been provided in Ref. \cite{shi2022dynamic},
we establish the conclusion from the perspective of pseudo-spin vectors
here. We rewrite the form of $F\left( k\right) $, 
\begin{equation}
	F\left( k\right) =\frac{1}{2}\sin 2\vartheta _{{k}},  \label{zeta}
\end{equation}
with $\vartheta _{{k}}=\sin ^{-1}\left[ \frac{\overline{r}{y}\left( k\right) 
}{\left\vert \overline{\mathbf{r}}\left( k\right) \right\vert }\right] $ and
then 
\begin{equation}
	\overline{\zeta }=\frac{1}{4\pi }\int_{0}^{\pi }\left\vert \sin 2\vartheta
	_{k}\right\vert dk.
\end{equation}
Here we introduce a new angle $\varphi _{k}$, which is equal to $\vartheta
_{k}$ when $\Delta =1,$ 
\begin{equation}
	\sin \varphi _{k}=\frac{\overline{r}_{y}\left( k,\Delta =1\right) }{
		\left\vert \overline{\mathbf{r}}\left( k,\Delta =1\right) \right\vert }.
\end{equation}
It can be easily proven that the relations between $\varphi _{k}$ and $k$
are 
\begin{eqnarray}
	\left\vert \mathbf{B}\left( k,\Delta =1\right) \right\vert \sin \varphi _{k}
	&=&\sin k,  \label{aboutk1} \\
	\left\vert \mathbf{B}\left( k,\Delta =1\right) \right\vert \cos \varphi _{k}
	&=&\cos k-\mu .  \label{aboutk2}
\end{eqnarray}
Furthermore, we can establish the relation between $\varphi _{k}$ and $%
\vartheta _{k}$ 
\begin{equation}
	\tan \vartheta _{k}=\Delta \tan \varphi _{k}.
\end{equation}
Based on the relations in Eqs. (\ref{aboutk1}) and (\ref{aboutk2}), we can
express $dk$\ by $d\varphi _{k}$\ and then substitute the result into Eq. (%
\ref{zeta}). Thus, we obtain, 
\begin{equation}
	\overline{\zeta }=\frac{1}{4\pi }\int_{0}^{\Theta }\frac{\left\vert \sin
		2\vartheta _{k}\right\vert \left\vert \mathbf{B}\left( k,\Delta =1\right)
		\right\vert }{\cos \left( k-\varphi _{k}\right) }d\varphi _{k}.
\end{equation}
Now let us consider the contributions of the pair of pseudo-spin vectors $%
\overline{\mathbf{r}}(k_{1})$ and $\overline{\mathbf{r}}(k_{2})$ with the
same value of $\sin \vartheta _{k}$. In the nontrivial region, as shown in
Fig. \ref{figS5}(c), we have 
\begin{equation}
	\vartheta _{k_{1}}=\pi -\vartheta _{k_{2}}.
\end{equation}
In the trivial region, as shown in Fig. \ref{figS5}(d),\ we have 
\begin{equation}
	\vartheta _{k_{1}}=\vartheta _{k_{2}}.
\end{equation}
It is easy to observe that regardless of whether the region is trivial or
nontrivial, we obtain the following relation, 
\begin{equation}
	\frac{\left\vert \mathbf{B}\left( k_{1},\Delta =1\right) \right\vert }{\cos
		\left( k_{1}-\varphi _{k_{1}}\right) }+\frac{\left\vert \mathbf{B}\left(
		k_{2},\Delta =1\right) \right\vert }{\cos \left( k_{2}-\varphi
		_{k_{2}}\right) }=2.
\end{equation}
Thus, the value of $\overline{\zeta }(\infty )$ can be expressed as 
\begin{equation}
	\overline{\zeta }=\frac{1}{2\pi }\int_{0}^{\Theta }\left\vert \sin
	2\vartheta _{k}\right\vert d\varphi _{k}.
\end{equation}
\ Again the initial state, $\left\vert \psi (0)\right\rangle
=\prod_{l}^{N}\left\vert 0\right\rangle _{N}$, represents the GS of the
pre-quenched topological trivial Hamiltonian in the coordinate space when $%
\mu \rightarrow \infty $. It is evident that when the post-quenched
Hamiltonian resides in the nontrivial region, the upper limit $\Theta =\pi
/2 $, allowing $\zeta (\infty )$\ to attain a value independent of $\mu $.
However, in the case of the post-quenched Hamiltonian being in the
topologically trivial region, $\Theta =\arcsin \frac{1}{\mu }$. One can
readily observe the nonanalytic behavior of $\overline{\zeta }$ when the system crosses the boundary.

\begin{figure}[tbh]
	\centering\includegraphics[width=0.9\textwidth]{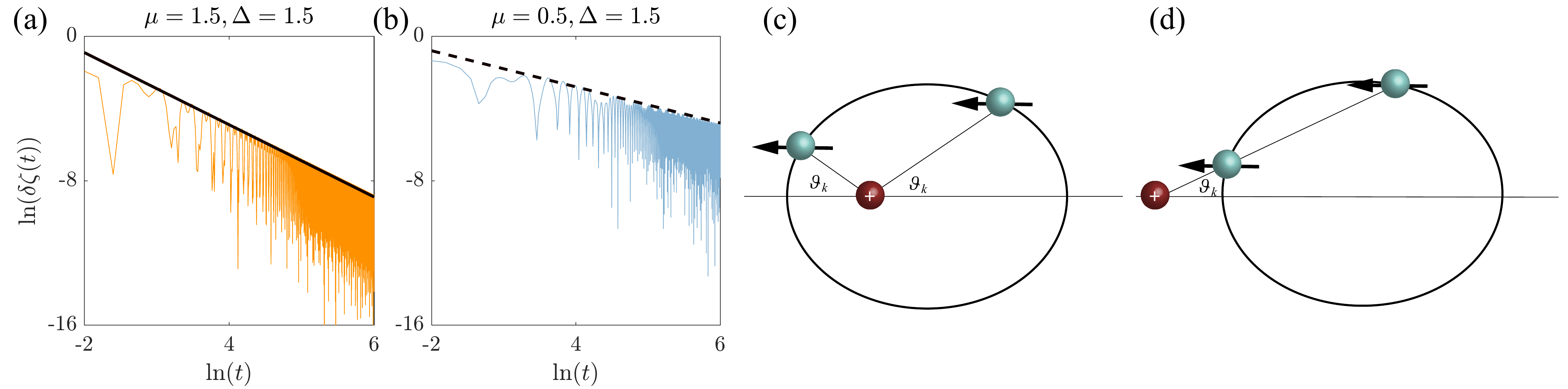} 
	\caption{Plots of $\ln (\protect\delta \protect\zeta (t))$\ as a function
		of $\ln (t)$. In (a) we set $\protect\mu =1.5$\ and $\Delta =0.5$ and the
		location of extrema $\left\vert \mathbf{B}\left( k\right) \right\vert $ is $		k_{0}=0.$ While in (b) $\protect\mu =0.5$\ and $\Delta=0.5$ and the location
		is $k_{0}=0.84$. The slopes of the solid black and dashed black lines in (a)
		and (b) are $-1$\ and $-0.5$, respectively. These results correspond with
		our analysis of the power law of $\protect\delta \protect\zeta (t)$. (c)-(d)
		sketched the pair of the pseudo spins, which serves as the building block
		for understanding the emergence of the plateau in Fig. 3(b) of the main text.
	} \label{figS5}
\end{figure}

\end{widetext}
\end{document}